\documentclass[
aps,
prb,
reprint,
superscriptaddress,
amsmath
]
{revtex4-2} 

\usepackage[utf8]{inputenc}
\usepackage{siunitx}
\usepackage{graphicx}
\usepackage{amssymb}
\usepackage{booktabs}
\usepackage{color,xcolor}
\usepackage{multirow}
\usepackage{braket}
\DeclareSIUnit\angstrom{\text {Å}}
\usepackage{hyperref}

\begin{document}

\title{Impact of strain on electron--phonon coupling of quantum emitters}

\author{Vytautas Žalandauskas}
\email{vytautas.zalandauskas@fys.uio.no}
\affiliation{Department of Physics/Centre for Materials Science and Nanotechnology, University of Oslo, 0316 Oslo, Norway}

\author{Rokas Silkinis}
\affiliation{Department of Fundamental Research, Center for Physical Sciences and Technology (FTMC), Vilnius LT-10257, Lithuania}

\author{Lukas Razinkovas}
\affiliation{Department of Fundamental Research, Center for Physical Sciences and Technology (FTMC), Vilnius LT-10257, Lithuania}

\author{Ali Tayefeh Younesi}
\affiliation{Max-Planck Institut für Polymerforschung, Ackermannweg 10, 55128 Mainz, Germany}

\author{Minh Tuan Luu}
\affiliation{Max-Planck Institut für Polymerforschung, Ackermannweg 10, 55128 Mainz, Germany}

\author{Ronald Ulbricht}
\affiliation{Max-Planck Institut für Polymerforschung, Ackermannweg 10, 55128 Mainz, Germany}

\author{Ulrike Grossner}
\affiliation{Advanced Power Semiconductor Laboratory, ETH Zürich, Physikstrasse 3, 8092 Zürich, Switzerland}

\author{Lasse Vines}
\affiliation{Department of Physics/Centre for Materials Science and Nanotechnology, University of Oslo, 0316 Oslo, Norway}

\author{Marianne Etzelm{\"u}ller Bathen}
\email{m.e.bathen@fys.uio.no}
\affiliation{Department of Physics/Centre for Materials Science and Nanotechnology, University of Oslo, 0316 Oslo, Norway}
\affiliation{Advanced Power Semiconductor Laboratory, ETH Zürich, Physikstrasse 3, 8092 Zürich, Switzerland}

\date{\today}

\begin{abstract}
Defects in semiconductors acting as optically active spin qubits are intriguing objects of fundamental study and future technological developments. These defect-based color centers are of particular interest for detection and response to physical variations such as pressure and strain, or conversely --- as we demonstrate the possibility of herein --- pressure and strain can be utilized to manipulate quantum emitter properties. To investigate how strain can alter the fundamental electron--phonon interaction of quantum defects, we employ the negatively charged silicon vacancy ($\mathrm{V_{Si}^{-}}$) in 4H-SiC as a use-case and study its vibrational structure under applied tensile and compressive uniaxial strain using first-principles calculations. We show that the strain variations of the emission spectrum can be explained by differing responses of bulk-like and quasi-localized vibrational modes. Importantly, the $\mathrm{V_{Si}^{-}}$ defect exhibits a strain-induced enhancement of the Debye--Waller factor under uniaxial tensile strain applied along the $a$-axis of 4H-SiC, thereby improving its performance as a quantum emitter. The strain-dependent changes in the phonon sideband enable distinguishing between compressive and tensile strain, opening up the possibility of magnetic-field-free strain detection using only spin-conserving transitions of solid-state quantum emitters.
\end{abstract}

\maketitle

\section{Introduction}

Defects in solids, such as color centers, are highly sensitive to their local environment, a property that underpins their broad use in quantum sensing applications~\cite{Degen_2017,Awschalom_2018}. External perturbations alter the electronic and spin degrees of freedom of these defects, leading to distinct changes in their optical properties that facilitate the detection of temperature, pressure, or electromagnetic fields~\cite{Toyli_2013,Kucsko_2013,Doherty_2014,Vindolet_2022,Hilberer_2023,Dolde_2011,Alaerts_2024, Rondin_2014}.
Among these perturbations, strain plays a unique role, as it directly alters the local crystal field and lattice dynamics, thereby affecting both optical transition energies and electron--phonon interactions~\cite{Meesala_2018,Bates_2021,Liu_2026,Mu_2026,Huang_2026,Hollendonner_2026}. These effects are observed as measurable shifts and intensity redistributions in the zero-phonon line (ZPL) and the phonon sideband (PSB) of defect emission. Recent advances in first-principles optical lineshape calculations now enable a quantitative description of color center spectral properties that allow defect identification and characterization through the vibrational fingerprints present in the PSB~\cite{Razinkovas_2021,Jin_2021,Jin_2022,Silkinis_2025,Linderalv_2025,Turiansky_2025}. However, a detailed microscopic understanding of defect--lattice interactions, with an emphasis on strain, that incorporates the intrinsic vibrational properties of solid-state color centers, is not yet available.

Among solid-state color centers, the nitrogen-vacancy (NV) center in diamond remains the most extensively studied and widely used system for quantum sensing applications~\cite{Doherty_2013}. Beyond diamond, silicon carbide (SiC) has emerged as an alternative platform, hosting a plethora of color centers with properties comparable to those of the NV center that exhibit environment-dependent spin and optical responses relevant for quantum sensing. In particular, emission from the silicon vacancy ($\mathrm{V_{Si}}$)~\cite{Widmann_2015} and divacancy ($\mathrm{V_{Si}V_{C}}$) defects~\cite{Christle_2015} in 4H-SiC has been shown to respond to electric fields through a measurable Stark effect~\cite{Casas_2017,Bathen_2019,Udvarhelyi_2020,Ruhl_2020,Anderson_2019}, while a pronounced strain response has been demonstrated for $\mathrm{V_{Si}}$ in 6H-SiC microparticles~\cite{Vasquez_2020}. A quantitative understanding of strain-induced spectral shifts is essential for utilizing these defects as local pressure or stress gauges and for determining whether controlled strain can be employed to tailor and enhance their optical properties. The 4H-SiC polytype hosts well-characterized color centers which are sensitive to external perturbations~\cite{Castelletto_2023}, making it an ideal platform for probing the interplay between electronic structure, vibrational structure, and external pressure or strain. Previous work~\cite{Udvarhelyi_2020} investigated the strain dependence of emission from the $\mathrm{V_{Si}}$ in 4H-SiC but did not consider strain-induced modifications of the vibrational structure. We therefore consider the negatively charged silicon vacancy ($\mathrm{V_{Si}^{-}}$) in 4H-SiC as a test case for developing a comprehensive theoretical framework for strained solid-state color centers.

In this work, we present emission lineshapes calculated using state-of-the-art methodology for $\mathrm{V_{Si}^{-}}$ under tensile and compressive uniaxial strains applied along the $a$-axis of 4H-SiC, consistent with the strain geometry used in experimental studies of strain-induced ZPL shifts in SiC micro- and nanoparticles~\cite{Vasquez_2020}, and compare these results with experimental emission spectra at zero strain. We find that the applied strain alters the electron--phonon coupling strength, enabling the tuning of quantum emitter properties. Indeed, we predict that the $\mathrm{V_{Si}^{-}}$ defect exhibits an enhanced Debye--Waller factor under $+2\%$ uniaxial strain along the $a$-axis as compared to the strain-free case, demonstrating that strain, or other external perturbations, can be employed to drive quantum emitters into a regime with improved properties. Finally, the electron--phonon interactions giving rise to the broad PSB are suggested as a potential readout mechanism capable of distinguishing between tensile and compressive strain.

\section{Methods}

\subsection{First-principles calculations}

The defect states were studied using density functional theory (DFT) calculations within the Kohn--Sham (KS) formalism. Computations were performed using the Vienna \textit{ab initio} Simulation Package (VASP)~\cite{Kresse_1993,Kresse_1994,Kresse_1996a}. The projector augmented-wave (PAW) method~\cite{Blochl_1994,Kresse_1996b} and plane waves were used to describe core and valence electrons, respectively. We utilized the meta-GGA class r$^2$SCAN functional~\cite{Furness_2020}. This functional was selected for its capability in accurately capturing the structural, electronic, and vibrational properties of deep-level defects in 4H-SiC~\cite{Abbas_2025,Zalandauskas_2025}. Hexagonal 4H-SiC supercells containing $400$ atoms were constructed by repeating the primitive cell $5\times5\times2$ times along the crystallographic axes. All calculations employed $\Gamma$-point sampling of the Brillouin zone and a plane-wave cutoff energy of $600$~eV. The electronic self-consistent field (SCF) cycle and atomic forces were converged to within $10^{-8}$~eV and \SI{5}{\milli\electronvolt\per\angstrom}, respectively. Defects were created by removing a silicon atom from the lattice to form a silicon vacancy ($\mathrm{V_{Si}}$). Calculations were performed on the negative charge state of the $\mathrm{V_{Si}}$ ($\mathrm{V_{Si}^{-}}$) which is the only known optically bright state of this defect center.

\subsection{Vibrational structure calculations}

Phonon modes for both bulk and defect supercells were calculated using the finite displacement method with configurations generated using the {PHONOPY} package~\cite{Togo_2015}. Vibrational modes were extrapolated to the dilute limit using the embedding methodology of Refs.~\cite{Alkauskas_2014,Razinkovas_2021}, with further details provided in the Supplemental Material Sec.~S3.

\subsection{Experimental setup}

To measure the emission spectra at zero strain, we employ transient absorption spectroscopy~\cite{Younesi_2022} on a $c$-cut 4H-SiC substrate containing ensembles of $\mathrm{V_{Si}}$~\cite{Younesi_2022,Luu_2024}. In contrast to traditional luminescence methods, where \textit{spontaneous} emission is measured, we here measure the \textit{stimulated} emission, which enables selective ZPL excitation and diminishes the spectral overlap between the emission of $\mathrm{V_{Si}^{-}}(h)$ and $\mathrm{V_{Si}^{-}}(k)$ defect ensembles. See the Supplemental Material Sec.~S6 for additional details and Ref.~\cite{Younesi_2026}.

\section{Results and discussion}

\subsection{Electronic properties}

\begin{figure}[t]
  \includegraphics[width=1\columnwidth]{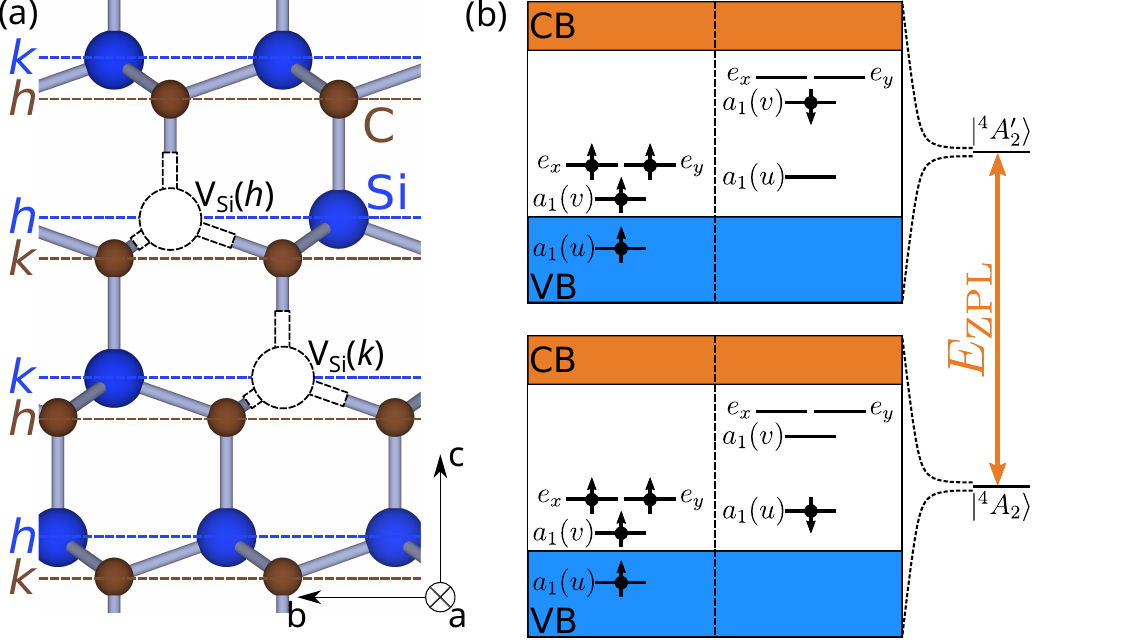}
  \caption{(a) Ball-and-stick representation of the silicon vacancy ($\mathrm{V_{Si}}$) defect in 4H-SiC. Silicon and carbon atoms are shown as blue and brown spheres, respectively, while the vacancy site is depicted as a hollow sphere. Two nonequivalent configurations, hexagonal $\mathrm{V_{Si}}(h)$ and quasicubic $\mathrm{V_{Si}}(k)$, are illustrated. The planes labeled \textit{h} and \textit{k} indicate the symmetry of the corresponding lattice sites.
  (b) Schematic representation of single-particle defect levels in the band gap of 4H-SiC showing their occupations in the ground $^{4}A_{2}$ state (bottom) and excited $^{4}A'_{2}$ state (top) of $\mathrm{V_{Si}^{-}}$. The spin-majority channel (left side) is denoted with upward arrows and the spin-minority channel (right side) with downward arrows. Shaded areas correspond to the valence band (VB) in blue and the conduction band (CB) in orange.}
  \label{fig:defects}
\end{figure}

\begin{table}[b]
    \centering
    \caption{Calculated ZPL shifts (in meV) for $\mathrm{V_{Si}^{-}}(h)$ and $\mathrm{V_{Si}^{-}}(k)$ under uniaxial strain along the $a$-axis of 4H-SiC and compared to experimental data.}
    \begin{tabular}{| c | c c | c |}
    \hline
    Strain & $\mathrm{V_{Si}^{-}}(h)$ & $\mathrm{V_{Si}^{-}}(k)$ & Expt.~\cite{Vasquez_2020}  \\
    \hline 
    $-2\%$ &  $26.2$ &  $35.2$ & $26$ ($\mathrm{V_{Si}^{-}}(h)$, 6H-SiC) \\ 
    $+2\%$ & $-30.9$ & $-43.7$ & \\
    \hline
    \end{tabular}
    \label{tab:zplshift}
\end{table}

In the 4H-SiC polytype, the Si--C bilayers stack along the crystallographic $c$-axis ([0001] direction), where each atomic site can adopt either a hexagonal (\textit{h}) or quasicubic (\textit{k}) configuration. Consequently, two nonequivalent configurations of the silicon vacancy, $\mathrm{V_{Si}}(h)$ and $\mathrm{V_{Si}}(k)$, must be considered [see Fig.~\ref{fig:defects}(a)]. The $\mathrm{V_{Si}^{-}}$ has a spin-quartet ground state with $A_{2}$ orbital symmetry ($^4\!A_{2}$), arising from five electrons localized on the four carbon dangling bonds surrounding the vacancy. The lowest-energy optical excitation corresponds to promoting one electron from the $a_{1}(u)$ orbital to $a_{1}(v)$ in the spin-minority channel [see Fig.~\ref{fig:defects}(b)], giving rise to an excited $^4\!A'_{2}$ state. Both states retain $C_{3v}$ symmetry in the unstrained crystal~\cite{Soykal_2016}, which gets reduced to $C_{1}$ under applied uniaxial strain along the $a$-axis.

The calculated ZPL energies for the $^4\!A'_{2} \rightarrow {^4\!A_{2}}$ optical transition obtained using the r$^{2}$SCAN functional are 1.393~eV for $\mathrm{V_{Si}^{-}}(h)$ and 1.307~eV for $\mathrm{V_{Si}^{-}}(k)$, which are in good agreement with the experimental V1 (1.439~eV) and V2 (1.353~eV) values, respectively. Application of uniaxial strain along the crystallographic $a$-axis of 4H-SiC results in substantial shifting of the emission energies for both configurations. The calculated ZPL energy shifts for $\mathrm{V_{Si}^{-}}(h)$ and $\mathrm{V_{Si}^{-}}(k)$ are contained in Table~\ref{tab:zplshift}. Notably, the calculated shift in $\mathrm{V_{Si}^{-}}(h)$ emission energy of 26.2~meV is in excellent agreement with experimental data, where a 26~meV ZPL shift was observed for the $\mathrm{V_{Si}^{-}}(h)$ defect in 6H-SiC microparticles and attributed to $-2\%$ compressive strain along the particle's $a$-axis~\cite{Vasquez_2020}, in support of the chosen methodology.

\subsection{Vibrational properties}

\begin{figure*}[t]
  \includegraphics[width=1\textwidth]{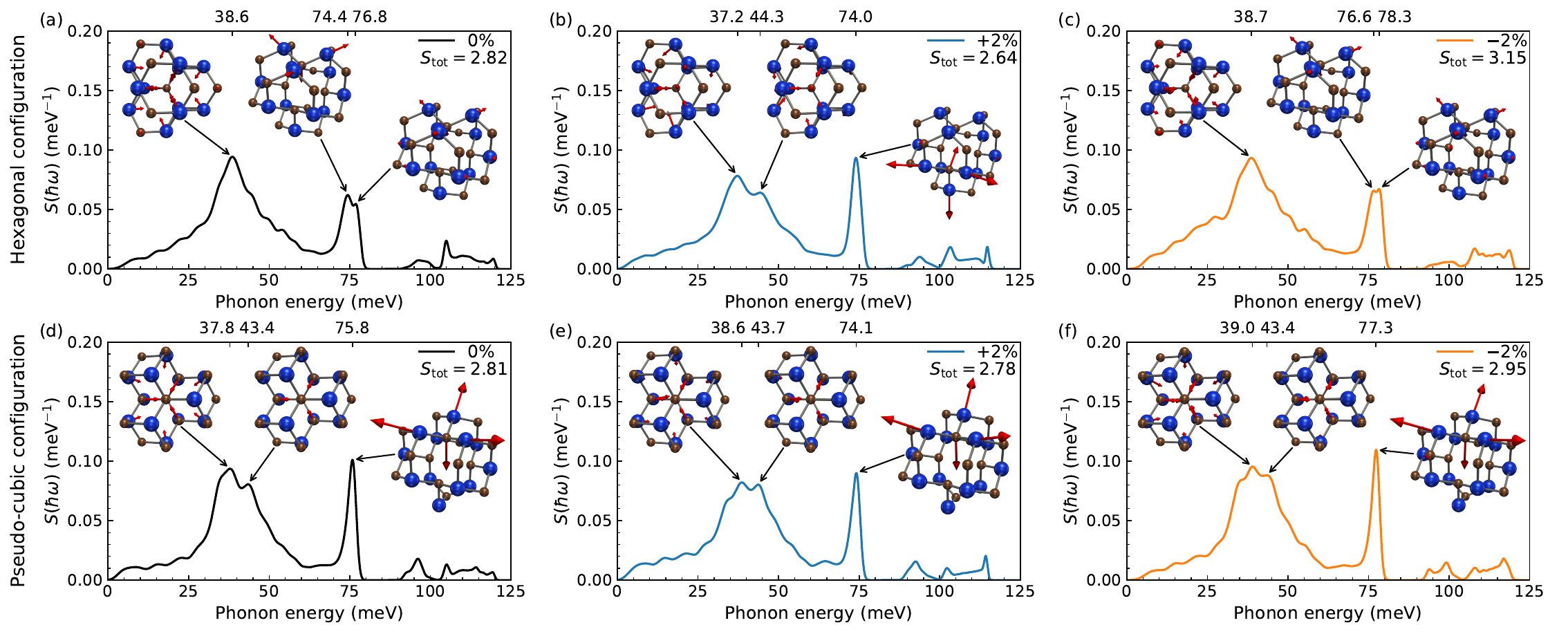}
  \caption{Spectral functions of electron--phonon coupling $S(\hbar\omega)$ (in meV$^{-1}$) for the emission of the $\mathrm{V_{Si}^{-}}$ defect in 4H-SiC under uniaxial strain applied along the $a$-axis. Panels (a)--(c) correspond to the hexagonal $\mathrm{V_{Si}^{-}}(h)$ configuration, while (d)--(f) show results for the quasicubic $\mathrm{V_{Si}^{-}}(k)$ configuration. Each column shows results for strain magnitudes of $0\%$, $+2\%$, and $-2\%$, respectively. Insets illustrate the effective shapes of collective vibrations corresponding to delocalized bulk-like and quasi-localized modes. Amplitudes are scaled for visual clarity.}
  \label{fig:modes_strain_grid}
\end{figure*}

Phonon sidebands in the emission spectrum arise from vibrational excitations following the optical transition. They are quantitatively described by the spectral function of electron--phonon coupling $S(\hbar\omega)$, which characterizes the average number of phonons emitted within a given energy interval. Figures~\ref{fig:modes_strain_grid}(a)--\ref{fig:modes_strain_grid}(f) present the spectral functions of electron--phonon coupling, together with corresponding shapes of collective vibrations that govern the emission of the $\mathrm{V_{Si}^{-}}$ defect in 4H-SiC under zero strain and uniaxial strain along the $a$-axis. Panels~(a)--(c) correspond to the hexagonal configuration $\mathrm{V_{Si}^{-}}(h)$, whereas panels~(d)--(f) display the results for the quasicubic configuration $\mathrm{V_{Si}^{-}}(k)$. The three columns represent strain-free ($0\%$, black curves), tensile ($+2\%$, blue curves), and compressive ($-2\%$, orange curves) conditions, respectively. Across both configurations, two vibrational features dictate the electron--phonon coupling: delocalized bulk-like modes in the range of 30--50~meV, and higher-energy quasi-localized modes within the 67--83~meV range. These quasi-local modes are associated with the motion of a neighboring carbon atom vibrating along the principal $c$-axis towards the silicon vacancy site and vibrations of three nearest neighbor silicon atoms. For the $\mathrm{V_{Si}^{-}}(h)$ configuration, the neighboring carbon atom is at the $h$ lattice site and its three nearest silicon neighbors are located at $k$ symmetry sites. In contrast, for $\mathrm{V_{Si}^{-}}(k)$, the vibrations are centered on the neighboring carbon atom at the $k$ lattice site and its adjacent silicon atoms at $h$ sites.

In the absence of strain, the electron--phonon coupling exhibits pronounced differences between the two defect configurations. For $\mathrm{V_{Si}^{-}}(h)$, the spectral function is dominated by a low-energy peak at approximately 38.6~meV, accompanied by a sharp split feature at 74.4 and 76.8~meV. In contrast, the $\mathrm{V_{Si}^{-}}(k)$ displays a distinctly different response, characterized by a double-peak structure at 37.8 and 43.4~meV, while exhibiting a single dominant sharp feature at approximately 75.8~meV. These distinctions in the electron--phonon coupling are also reflected in the experimental emission spectra discussed further below.

Under uniaxial strain applied along the $a$-axis, the vibrational structure evolves as follows. The bulk-like vibrational modes show marginal variations in their electron--phonon coupling peak positions, indicating weak sensitivity to strain. This is consistent with the negligible changes in the calculated bulk phonon density of states (DOS) in the acoustic regime (see Fig.~S3 of the Supplemental Material). The quasi-local modes shift to lower energies under $+2\%$ tensile strain, whereas under $-2\%$ compressive strain they shift toward higher energies. A notable exception occurs for $\mathrm{V_{Si}^{-}}(h)$ under $+2\%$ strain [see Fig.~\ref{fig:modes_strain_grid}(b)], where the vibration of the neighboring carbon atom no longer occurs along the $c$-axis but instead involves the carbon atom at the $k$ site and three neighboring silicon atoms (two at $h$ sites and one at a $k$ site). This strain-induced modification renders the hexagonal configuration more similar to the quasicubic configuration, leading to a splitting of the bulk-like electron--phonon coupling feature and a single quasi-local vibrational peak.

The calculated total Huang--Rhys (HR) factor~\cite{Huang_1950}, defined as $S_\mathrm{tot}=\sum_{k} S_{k}$, quantifies the overall strength of electron--phonon coupling through the cumulative contribution of all vibrational modes $k$. For the hexagonal configuration, the $S_{\mathrm{tot}}$ were determined to be $2.82$, $2.64$, and $3.15$ under $0\%$, $+2\%$, and $-2\%$ uniaxial strain along the $a$-axis, respectively. In the quasicubic configuration, the $S_{\mathrm{tot}}$ were $2.81$, $2.78$, and $2.95$ for the respective strain conditions (see also legends in Fig.~\ref{fig:modes_strain_grid}). The calculated mass-weighted displacement $\Delta Q$ ($\mathrm{amu}^{0.5}$\,\r{A}) between the ground- and excited-state equilibrium geometries for $\mathrm{V_{Si}^{-}}(h)$ was $0.706$ at zero strain, decreasing to $0.642$ under $+2\%$ strain, and increasing to $0.718$ under $-2\%$ strain. In contrast, $\mathrm{V_{Si}^{-}}(k)$ exhibits a weaker strain dependence, with $\Delta Q$ values of $0.696$, $0.680$, and $0.705$ for $0\%$, $+2\%$, and $-2\%$ strain, respectively. Tensile strain reduces the coupling to defect modes by increasing ion--ion separations around the vacancy site, while simultaneously enhancing coupling to low-energy bulk-like vibrational modes. For $\mathrm{V_{Si}^{-}}(k)$, the weaker strain dependence of $\Delta Q$ results in a marginal decrease of $S_{\mathrm{tot}}$ under $+2\%$ strain. In contrast, $\mathrm{V_{Si}^{-}}(h)$ exhibits stronger change of $\Delta Q$, leading to a larger reduction of $S_{\mathrm{tot}}$ under $+2\%$ strain.

\begin{figure*}
  \includegraphics[width=\textwidth]{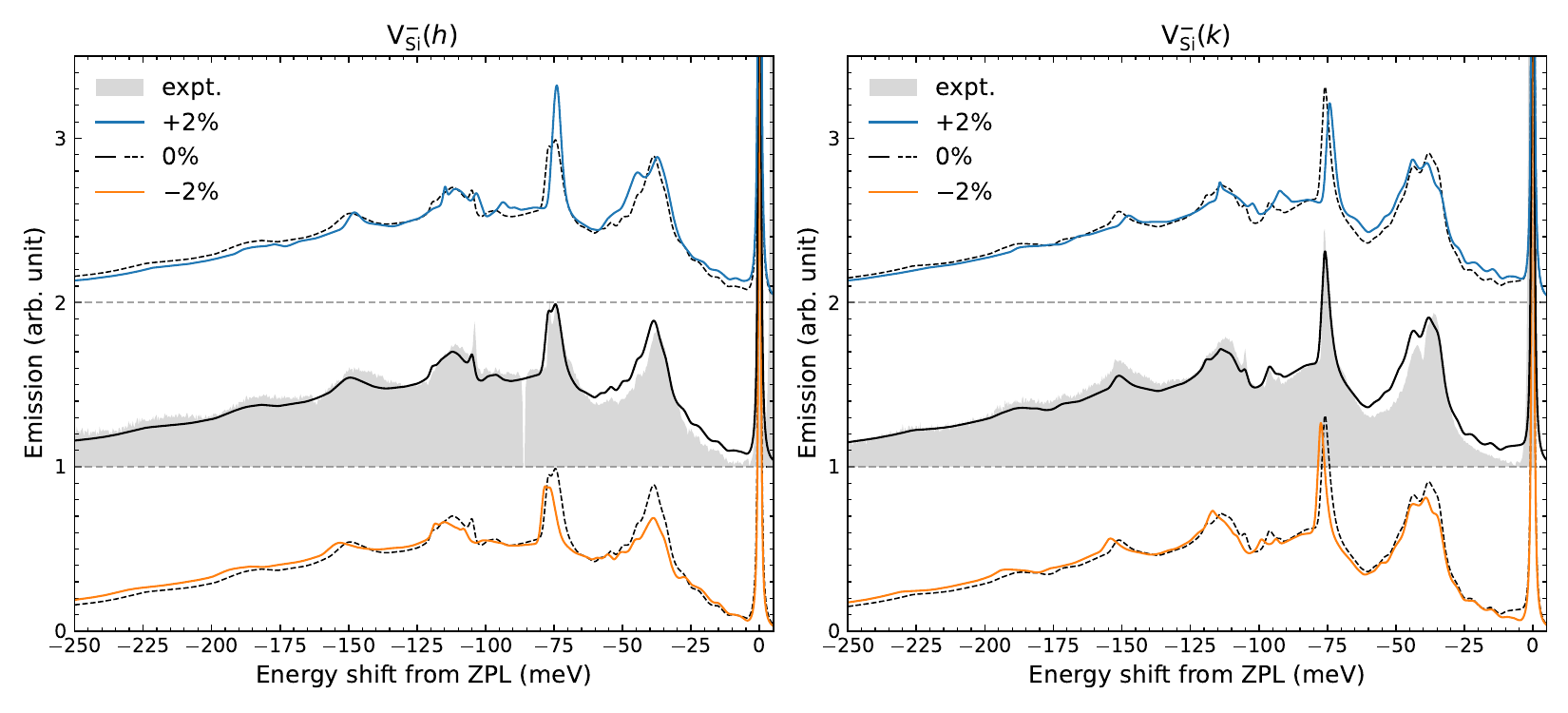}
  \caption{Calculated emission lineshapes for supercells with $\mathrm{V_{Si}^{-}}(h)$ and $\mathrm{V_{Si}^{-}}(k)$ defects with zero (middle panels, black curves), $+2\%$ (top panels, blue curves) and $-2\%$ (bottom panels, orange curves) uniaxial strain applied along $a$-axis. The shaded gray areas in the middle panels represent experimental data measured on $\mathrm{V_{Si}^{-}}$ ensembles for the strain-free case. Calculated strain-free lineshapes are also shown as black dotted lines in the top and bottom panels, allowing comparison with the calculated strained lineshapes. Emission lineshapes were computed with the r$^{2}$SCAN functional using HR theory and extrapolated to the dilute limit, approximated by a $25\times25\times8$ supercell with 40\,000 atomic sites.}
  \label{fig:lum_abs_strain}
\end{figure*}

\subsection{Optical lineshapes under strain}

Figure~\ref{fig:lum_abs_strain} shows calculated emission lineshapes for $\mathrm{V_{Si}^{-}}(h)$ and $\mathrm{V_{Si}^{-}}(k)$ defects under zero strain conditions and uniaxial strain applied along the $a$-axis. The curves correspond to strain-free ($0\%$, black curves), tensile ($+2\%$, blue curves), and compressive ($-2\%$, orange curves) conditions. The PSB observed in all spectra originates from bulk-like, quasi-local and localized vibrational modes and their respective replicas, with the PSB governed by the bulk-like and quasi-local contributions. All PSB peak positions will be given in units of meV below the ZPL. Strain-induced shifts of characteristic emission peaks can be identified in the PSB, reflecting the varying strain sensitivity of different phonon modes.

At zero strain, the calculated lineshapes (black lines) closely follow the experimental data (gray shaded areas in Fig.~\ref{fig:lum_abs_strain}). In particular, they reproduce the characteristic double-peak of the quasi-local vibrational modes for $\mathrm{V_{Si}^{-}}(h)$, observed at $(74.2,\,76.5)$~meV, as well as the double-peak of bulk-like vibrational modes for $\mathrm{V_{Si}^{-}}(k)$ at $(36.2,\,42.0)$~meV. This demonstrates that ensemble measurements based on stimulated emission can be directly compared with theoretically computed lineshapes when the ZPLs of different defect configurations are well separated and do not overlap with prominent vibrational features of other configurations.

For $\mathrm{V_{Si}^{-}}(h)$, the calculated unstrained emission spectrum features a prominent PSB peak at 38.6~meV arising from bulk-like vibrational modes, as well as a double-peak at $(74.5,\,76.8)$~meV associated with quasi-local vibrational modes. Under $-2\%$ compressive strain, these features shift to higher energies, appearing at 38.8~meV and $(76.8,\,78.2)$~meV. Under $+2\%$ tensile strain, the bulk-like peak splits into a double peak of $(37.3,\,44.5)$~meV, while the quasi-local mode double-peak collapses into a single peak at 74.0~meV, due to increased resonance with the bulk phonon modes. A faint feature corresponding to the phonon replica of the quasi-local mode is calculated to be at 149.5, 147.8, and 153.2~meV for $0\%$, $+2\%$, and $-2\%$ strain, respectively.

In contrast, $\mathrm{V_{Si}^{-}}(k)$ exhibits a double-peak structure in the bulk-like vibrational region, with features calculated at $(38.1,\,43.6)$~meV, while the quasi-local vibrational mode appears as a single peak at 75.8~meV. Under $+2\%$ tensile strain, these features shift to $(38.8,\,43.9)$~meV and 74.2~meV, whereas under $-2\%$ compressive strain they move to higher energies of $(39.1,\,43.7)$~meV and 77.4~meV. The quasi-local phonon replica is calculated to be at 151.5, 147.6, and 154.2~meV for $0\%$, $+2\%$, and $-2\%$ strain, respectively.

The first PSB peak associated with bulk-like vibrational modes shows minor energy shifts under strain consistent with changes in the bulk phonon DOS (see Fig.~S3 of the Supplemental Material). Stronger shifts are predicted for the second PSB peak associated with quasi-local vibrational modes. In both $\mathrm{V_{Si}^{-}}$ configurations, the second PSB peak shifts closer to the ZPL under tensile ($+2\%$) strain and away from it under compressive ($-2\%$) strain. Finally, the PSB feature near 150~meV, arising from the phonon replica of quasi-local modes, undergoes the largest strain-induced shifts among all observed peaks.

The Debye--Waller factor (DWF) is defined as the ratio of the ZPL intensity to the total emission lineshape and is an important quantum emitter characteristic. The calculated DWF values for $\mathrm{V_{Si}^{-}}$ defects under different strain conditions are presented in Table~\ref{tab:dq_hrf_dwf_strain}. For the hexagonal configuration, the calculated DWF values are $8.02\%$, $9.36\%$, and $5.96\%$ at $0\%$, $+2\%$, and $-2\%$ strain, respectively. Thus, strain can be utilized to enhance the DWF, as exemplified by the $+2\%$ condition. In the quasicubic configuration, the DWF values are $8.20\%$, $8.32\%$, and $7.32\%$ for the same strain values. These results are consistent with the HR factor trends, as stronger electron--phonon coupling leads to a reduction in the ZPL intensity and, consequently, a lower DWF.

\begin{table}
    \centering
    \caption{Computed mass-weighted displacements ($\Delta Q$, in $\mathrm{amu}^{0.5}$\,\r{A}), total Huang--Rhys factors ($S_\mathrm{tot}$), and Debye--Waller factors (DWF, in \%) for the emission of $\mathrm{V_{Si}^{-}}$ defects in 4H-SiC under $\pm 2$\% uniaxial strain along the $a$-axis.}
    \label{tab:dq_hrf_dwf_strain}
    \vspace{4pt}
    \begin{tabular}{| c | c c c | c c c |}
        \hline
        \multirow{2}{*}{Strain} 
        & \multicolumn{3}{c|}{$\mathrm{V_{Si}^{-}}(h)$} 
        & \multicolumn{3}{c|}{$\mathrm{V_{Si}^{-}}(k)$} \\
        \cline{2-7}
        & $\Delta Q$ & $S_\mathrm{tot}$ & DWF 
        & $\Delta Q$ & $S_\mathrm{tot}$ & DWF \\
        \hline
        $0\%$  
        & 0.706 & 2.82 & 8.02 
        & 0.696 & 2.81 & 8.20 \\
        $+2\%$ 
        & 0.642 & 2.64 & 9.36 
        & 0.680 & 2.78 & 8.32 \\
        $-2\%$ 
        & 0.718 & 3.15 & 5.96 
        & 0.705 & 2.95 & 7.32 \\
        \hline
    \end{tabular}
\end{table}

Finite-temperature effects on the emission spectra are presented in Supplemental Material Sec.~S5. The key PSB features remain clearly resolved up to liquid-nitrogen temperatures, while they become effectively smeared out at room temperature. Consequently, the PSB peak associated with quasi-local phonon modes, together with its replica, emerges as a robust spectral fingerprint for strain calibration in $\mathrm{V_{Si}^{-}}$ centers. Notably, the $\mathrm{V_{Si}^{-}}(k)$ configuration, which can be selectively addressed and probed using standard spontaneous emission spectroscopy, exhibits comparatively sharper PSB features than the $\mathrm{V_{Si}^{-}}(h)$ configuration, rendering it more favorable for strain studies.

\section{Conclusions}

We have presented a first-principles investigation of the vibrational structure and emission lineshapes of $\mathrm{V_{Si}^{-}}$ centers in 4H-SiC under uniaxial tensile and compressive strain applied along the $a$-axis. At zero strain, the calculated emission lineshapes are in excellent agreement with experimental spectra, reproducing the characteristic double-peak features associated with quasi-local modes for $\mathrm{V_{Si}^{-}}(h)$ and bulk-like modes for $\mathrm{V_{Si}^{-}}(k)$, validating the theoretical framework. The calculated emission spectra show that bulk-like modes are weakly affected by strain, whereas quasi-local, quasi-local replica, and localized modes exhibit clear, directional energy shifts under tensile and compressive strain. The emission lineshapes were modeled at 0~K, where fine features of the PSB are resolved. By modeling finite-temperature effects we demonstrate that the key PSB features remain clearly resolved up to liquid-nitrogen temperatures, while they become effectively smeared out at room temperature. Thus, overall PSB energy shifts remain a clear indicator of whether the defect experiences tensile or compressive strain along the $a$-axis. Taken together, our results establish that strain-dependent phonon sidebands, combined with ZPL shifts, provide multidimensional and magnetic-field-free access to strain using spin-conserving optical transitions. A comprehensive investigation of strain applied along different crystallographic directions remains an important direction for future work.

Beyond the observed spectral shifts in both ZPL and PSB, we find that uniaxial strain can beneficially impact the electron--phonon coupling. For the hexagonal $\mathrm{V_{Si}^{-}}(h)$ configuration, $+2\%$ tensile strain along the $a$-axis leads to a noticeable reduction of the overall electron--phonon coupling strength and an increased DWF. For the $\mathrm{V_{Si}^{-}}(k)$ configuration, on the other hand, the corresponding changes are marginal. Overall, these results demonstrate that strain can be employed as an external means to enhance quantum emitter performance.

Finally, this work demonstrates a theoretical framework for understanding the interaction between strain and spin-conserving optical transitions of solid-state quantum emitters, focusing on the case study of the V$_\mathrm{Si}^-$ in 4H-SiC. Importantly, other defects, either in SiC or alternative host materials, may possess more strongly localized vibrational modes that give rise to sharper PSB features and larger strain-induced spectral modifications.

\begin{acknowledgments}
The work of MEB was supported by an ETH Zürich Postdoctoral Fellowship. Financial support was kindly provided by the Research Council of Norway and the University of Oslo through the frontier research projects QuTe (no. 325573, FriPro ToppForsk-program) and TASQ (no. 354419, FriPro), and through the Centre for Defects in Semiconductors for Quantum Sensing (no.~354831). 
RU acknowledges funding by the Max-Planck Society. The computations were performed on resources provided by UNINETT Sigma2 --- the National Infrastructure for High Performance Computing and Data Storage in Norway, supercomputer GALAX of the Center for Physical Sciences and Technology, Lithuania, and the High Performance Computing Center “HPC Saulėtekis” in the Faculty of Physics, Vilnius University.
\end{acknowledgments}

\section*{Author contributions}
MEB designed the research with support from LR, UG and LV. VŽ, RS and MEB performed the calculations. VŽ, RS, and LR wrote the codes for lineshape calculations and embedding methodology. VŽ, RS, LR and MEB analyzed the data. RU supervised the experimental part. ATY and MTL performed the experiments. All authors have contributed to the discussion and interpretation of the results. VŽ wrote the manuscript with support and editing from all authors.

\bibliography{ref}

\clearpage
\onecolumngrid

\section*{Supplemental Material}

\makeatletter
\setcounter{secnumdepth}{3}
\def\@seccntformat#1{\csname the#1\endcsname\quad} 
\makeatother

\setcounter{section}{0}
\renewcommand{\thesection}{S\arabic{section}.}
\renewcommand{\thesubsection}{\Alph{subsection}.}

\setcounter{figure}{0}
\renewcommand{\thefigure}{S\arabic{figure}}

\setcounter{table}{0}
\renewcommand{\thetable}{S\arabic{table}}

\setcounter{equation}{0}
\renewcommand{\theequation}{S\arabic{equation}}

\section{Electronic structure and zero-phonon line energies}

The electronic structure of both silicon vacancy configurations in 4H-SiC can be understood in terms of four carbon dangling bonds surrounding the vacant silicon site~\cite{Soykal_2016}. These combine to form four molecular orbitals: two $a_{1}$ states [a lower-lying $a_{1}(u)$, a higher-lying $a_{1}(v)$] state, and a pair of degenerate orbitals with $e$ symmetry within the band gap (see Fig.~\ref{fig:defects}(b) from main text). In the negatively charged state that exhibits bright emission, five electrons occupy these orbitals. The ground state of the defect is a spin-quartet with $A_{2}$ orbital symmetry ($^{4}\!A_{2}$), described by a single-determinant wave function $|a_{1}(u)\,\bar{a}_{1}(u)\,a_{1}(v)\,e_{x}\,e_{y}|$ for the spin projection $m_{s}=3/2$. The first excited state $^4\!A'_{2}$ can be represented by a single-determinant wave function $|a_{1}(u)\,a_{1}(v)\,\bar{a}_{1}(v)\,e_{x}\,e_{y}|$ for the same spin projection. Both states exhibit $C_{3v}$ point-group symmetry characterized by a threefold rotation axis $C_{3}$ oriented along the crystallographic $c$-axis, and three vertical mirror planes $\sigma_v$. When uniaxial strain is applied along the $a$-axis, the threefold rotational symmetry and all mirror symmetries are lifted. As a result, the symmetry is reduced to the trivial point group $C_{1}$, and all orbitals transform according to the totally symmetric representation $a$.

The ZPL energies were computed using the $\Delta$-self-consistent-field ($\Delta$SCF) approach~\cite{Gali_2009}, whose reliability has been demonstrated for a variety of point defects in 4H-SiC~\cite{Jin_2021,Zhu_2021,Abbas_2025,Zalandauskas_2025}. The excited state was constructed by constraining the Kohn--Sham single-particle level occupancies and promoting one electron from the highest occupied level in the spin-minority channel to the next higher-energy level within the same spin channel. In the zero strain case, this corresponds to a promotion of an electron from the $a_{1}(u)$ defect level to the nearest higher-lying $a_{1}(v)$ defect level in the spin-minority channel. Under uniaxial strain, the excitation is modeled by promoting an electron from the highest occupied $a$-symmetry defect orbital to the nearest higher-lying unoccupied $a$-symmetry defect level within the spin-minority channel.

Table~\ref{tab:ZPL_SM} presents a comparison between the calculated and experimental ZPL energies for the $\mathrm{V_{Si}^{-}}(h)$ and $\mathrm{V_{Si}^{-}}(k)$ defects in 4H-SiC. Previously reported theoretical ZPL values are also included, along with brief computational details provided below the table. 
The meta-GGA r$^2$SCAN functional underestimates the ZPL energies by 0.046~eV for both $\mathrm{V_{Si}^{-}}(h)$ and $\mathrm{V_{Si}^{-}}(k)$ defects. For comparison, previous theoretical studies employing the PBE functional report larger underestimations, ranging from 0.166 to 0.194~eV for $\mathrm{V_{Si}^{-}}(h)$ and from 0.155 to 0.178~eV for $\mathrm{V_{Si}^{-}}(k)$~\cite{Csore_2021,Davidsson_2022}. Despite the fact that r$^2$SCAN underestimates the bulk 4H-SiC band gap (2.223~eV for PBE, 2.611~eV for r$^2$SCAN, 3.172~eV for HSE06, and 3.265~eV from experiment~\cite{Choyke_1997}), it nevertheless yields ZPL energies that come close to the experimental values, while requiring $\approx 40$ times less computational effort than the hybrid HSE06 functional for the same supercell size and Brillouin zone sampling.

\begin{table}[b]
\centering
\caption{Comparison of calculated and experimental zero-phonon line (ZPL) energies (in units of eV) for spin-conserving transitions for the negatively charged silicon vacancy centers in 4H-SiC. Previous theoretically predicted ZPL energies are also shown.}
\label{tab:ZPL_SM}
\begin{tabular}{| c | c | c | c | c |}
\hline
& This work &  \multicolumn{2}{c|}{Previous theoretical work} & Expt. \\
\hline
& r$^2$SCAN & PBE & HSE06 & \\
\hline
$\mathrm{V_{Si}^{-}}(h)$ & 1.393 & 1.273$^{\mathrm{a}}$, 1.245$^{\mathrm{b}}$ &
1.450$^{\mathrm{a}}$, 1.376$^{\mathrm{b}}$, 1.541$^{\mathrm{c}}$, 1.450$^{\mathrm{d}}$, 1.57$^{\mathrm{e}}$ & 1.439 \\
\hline
$\mathrm{V_{Si}^{-}}(k)$ & 1.307 & 1.198$^{\mathrm{a}}$, 1.175$^{\mathrm{b}}$ & 1.385$^{\mathrm{a}}$, 1.283$^{\mathrm{b}}$, 1.443$^{\mathrm{c}}$, 1.385$^{\mathrm{d}}$, 1.24$^{\mathrm{e}}$, 1.35$^{\mathrm{f}}$ & 1.353 \\
\hline
\end{tabular}
\begin{flushleft}
$^{\mathrm{a}}$Reference~\cite{Csore_2021}: Calculations were carried out using VASP code with PAW pseudopotentials (PPs). The plane-wave energy cutoff was set to 420~eV. The Brillouin zone was sampled with the $\Gamma$ point. 1536 atom supercell was used. The ZPL values with the HSE06 functional were obtained by correcting the corresponding PBE result with the difference of the isolated $\mathrm{V_{Si}^{-}}$-related PBE and HSE06 ZPL values as an estimate.\\
$^{\mathrm{b}}$Reference~\cite{Davidsson_2022}: Calculations were carried out using VASP code with PAW PPs. The plane-wave energy cutoff was set to 600~eV for PBE and 420~eV for HSE06 functional. The Brillouin zone was sampled with the $\Gamma$ point. 2304 atom supercell was used for PBE functional while 576 atom supercell was used for HSE06.\\
$^{\mathrm{c}}$Reference~\cite{Ivady_2017}: Calculations were carried out using VASP code with PAW PPs. The plane-wave energy cutoff was set to 390~eV. The Brillouin zone was sampled with the $\Gamma$ point. 1536 atom supercell was used.\\
$^{\mathrm{d}}$Reference~\cite{Udvarhelyi_2020}: Calculations were carried out using the VASP code with PAW PPs. The plane-wave energy cutoff was set to 420~eV. The Brillouin zone was sampled with the $\Gamma$ point. 768 atom supercell was used.\\
$^{\mathrm{e}}$Reference~\cite{Hashemi_2021}: Calculations were carried out using VASP code with PAW PPs. The plane-wave energy cutoff was set to 400~eV. The Brillouin zone was sampled with the $\Gamma$ point. 400 atom supercell was used.\\
$^{\mathrm{f}}$Reference~\cite{Shang_2020}: Calculations were carried out using VASP code with PAW PPs. The plane-wave energy cutoff was set to 450~eV. The Brillouin zone was sampled using a $2\times2\times2$ $k$-point mesh. 400 atom supercell was used.\\
\end{flushleft}
\end{table}

\section{Theoretical description of electron--phonon coupling}

In the the Franck--Condon approximation, the normalized lineshape for emission and absorption at absolute zero temperature ($T = 0$~K) can be expressed as follows~\cite{Razinkovas_2021}:
\begin{equation}
  \label{eq:lineshape}
  L(\hbar\omega) = C \omega^{\kappa} A(\hbar\omega),
\end{equation}
where $C$ denotes a normalization constant, $A(\hbar\omega)$ denotes the phonon spectral function, and the exponent $\kappa$ is equal to 3 for emission and 1 for absorption. The spectral function is defined by the equation
\begin{equation}
  \label{eq:non_degen_spectral_function}
  A(\hbar\omega) = \sum_{m}
    \left| \braket{\chi_{i;0}|\chi_{f;m}} \right|^2
    \delta \left( E_{\mathrm{ZPL}} \mp (\varepsilon_{fm} - \varepsilon_{f0}) - \hbar\omega \right),
\end{equation}
where $E_{\mathrm{ZPL}}$ is the energy of the zero-phonon line (ZPL), while $\chi_{i;0}$ and $\chi_{f;m}$ represent the vibrational wave functions corresponding to the initial and final electronic states. Here, $\varepsilon_{fm}$ refers to the energy of the $m$-th vibrational state within the final electronic manifold relative to the potential energy minimum. The signs in front of the $\delta$-function argument differentiate between emission (minus sign) and absorption (plus sign). The optical spectral function $A(\hbar\omega)$ quantifies the transition amplitudes among vibrational states and plays a crucial role in determining the lineshape.

Due to the inherent differences in vibrational structures between the ground and excited states, the overlap integrals $\braket{\chi_{i;0}|\chi_{f;m}}$ found in Eq.~\eqref{eq:non_degen_spectral_function} are multidimensional. Given the complexity with numerous vibrational modes in defect systems, directly calculating these integrals proves computationally challenging. To circumvent these difficulties and streamline the evaluation of overlap integrals, we utilize the equal-mode approximation from Ref.~\cite{Razinkovas_2021,Markham_1959}, which assumes that both the shapes and frequencies of vibrational modes in the initial and final states are equivalent. To mitigate the drawbacks of this approximation, we consistently adopt the vibrational modes of the final state in our lineshape calculations, following guidelines from Ref.~\cite{Razinkovas_2021}.

The spectral function $A(\hbar\omega)$ can be effectively derived by calculating it in the time domain via the generating function approach established by Kubo and Lax~\cite{Kubo_1955,Lax_1952}. In this framework, $A(\hbar\omega)$ is extracted from the generating function $G(t)$ according to the following expression:
\begin{equation}
  \label{eq:opt_spectral_func_gen}
  A(\hbar\omega) = \frac{1}{2 \pi} \int_{-\infty}^{\infty}{G(t) e^{-\gamma |t|} e^{- i (E_{\mathrm{ZPL}} / \hbar - \omega) t} \mathrm{d}t},
\end{equation}
where the term $e^{-\gamma |t|}$ serves as a phenomenological adjustment to incorporate the homogeneous Lorentzian broadening of the ZPL that is not captured by this theoretical framework, in addition to addressing the effects of inhomogeneous broadening. The parameter $\gamma$ is fine-tuned to align with the experimentally observed linewidth of the ZPL. In all lineshape calculations the $\gamma$ value was set to be 0.30~meV.

Under the equal mode approximation, the expression for the generating function $G(t)$ can be represented as
\begin{equation}
  \label{eq:gen_func_with_sum}
  G(t) = \exp \left[ -S_{\mathrm{tot}} + \sum_{k} S_{k} e^{\pm i \omega_{k} t} \right],
\end{equation}
where the plus sign is applicable for emission and the minus sign for absorption, with $S_k$ denoting the partial Huang--Rhys (HR) factor, and the summation encompasses all vibrational modes within the system. The total coupling strength is denoted by $S_{\mathrm{tot}} = \sum_{k} S_{k}$. The factor $S_k$ quantifies the average number of excited phonons present during an optical transition~\cite{Huang_1950}, and is formulated as
\begin{equation}
  \label{eq:part_hr_factor}
  S_k = \frac{\omega_k \Delta Q_{k}^{2}}{2 \hbar},
\end{equation}
where $\Delta Q_k$ is the ionic displacement along the $k$-th normal mode, prompted by an optical transition. More specifically, $\Delta Q_k$ represents the projection of mass-weighted displacement between the ground and excited states onto the normalized phonon mode $\boldsymbol{\eta}_{k}$:
\begin{equation}
  \label{eq:delta_q}
  \Delta Q_{k} = \sum_{m\alpha}{\sqrt{M_{m}} \Delta R_{m\alpha} \eta_{k; m\alpha}}.
\end{equation}
In this equation, $\Delta R_{m\alpha}$ indicates the displacement of atom $m$ along the $\alpha$ coordinate, while $M_{m}$ is its corresponding mass.

To adequately describe the generating function in Eq.~\eqref{eq:gen_func_with_sum} in larger systems, where one would expect a continuum of vibrational frequencies, we introduce the spectral density of electron--phonon coupling, also referred to as the spectral function of electron--phonon coupling:
\begin{equation}
  \label{eq:spectral_function_of_ep_coupling}
  S(\hbar\omega) \equiv \sum_{k} S_{k} \delta(\hbar\omega_{k} - \hbar\omega),
\end{equation}
which enables us to rewrite the generating function in the following form:
\begin{equation}
  \label{eq:gen_func_with_integral}
  G(t) = \exp \left[-S_{\mathrm{tot}} + \int S(\hbar\omega) e^{\pm i \omega t} \text{d}{\omega} \right].
\end{equation}
We used a Gaussian smoothing function to facilitate the approximation of the $\delta$-functions in Eq.~\eqref{eq:spectral_function_of_ep_coupling}. This results in a smooth representation of the electron--phonon spectral function, where the $\delta$-functions are approximated using Gaussian functions with frequency-dependent widths $\sigma$ that gradually decrease from 1.5~meV at zero frequency to 0.5~meV at the maximum phonon frequency for emission.

Phonon modes of 4H-SiC supercells containing the $\mathrm{V_{Si}^{-}}$ defects were calculated using the finite-displacement approach, with atomic configurations generated by the {PHONOPY} package~\cite{Togo_2015}. A displacement amplitude of 0.01~\r{A} from the equilibrium geometry was employed. The relatively small size of the computationally tractable supercells employed in this study results in the absence of low-frequency acoustic modes and a poor description of vibrational resonances. Experimentally measured ensemble defect concentrations typically range from \SIrange{1e12}{1e16}{\per\centi\meter\cubed}. The phonon modes were extrapolated to the dilute-defect limit using the force-constant embedding methodology described in Sec.~\ref{sec:sm_embedding}.

An added challenge in supercell calculations occurs when determining the relaxation profile after an optical transition. The limited size of the supercell in direct calculations restricts the long-wavelength components of $\Delta R_{m\alpha}$ in Eq.~\eqref{eq:delta_q}. To accurately represent the relaxation profile in the dilute limit, we compute $\Delta Q_{k}$ for each vibrational mode of a larger supercell (obtained through the embedding procedure) by employing forces and the harmonic relation to displacements. The relaxation component $\Delta Q_{k}$ is calculated as follows:
\begin{equation}
  \label{eq:delta_Q_forces}
  \Delta Q_{k}
  = \frac{1}{\omega_{k}^{2}} \sum_{m\alpha}\frac{F_{m\alpha}}{\sqrt{M_{m}}} \eta_{k; m\alpha},
\end{equation}
where $F_{m\alpha}$ denotes the force acting on atom $m$ in the direction $\alpha$ when the system is in the final electronic state, yet retains the equilibrium geometry of the initial state calculated in the directly accessible supercell. Here $\eta_{k; m\alpha}$ is the vibrational mode shape as derived from the embedded supercell. By using forces that have already converged within the computationally tractable supercell, this strategy, in conjunction with the embedding methodology, effectively captures the electron--phonon coupling for low-frequency modes and offers an accurate description of the optical lineshapes.

In our calculations, the vibrational frequencies show very good agreement with experiment. However, with the r$^2$SCAN functional we observed an underestimation of atomic relaxations. We assume that the shape of the calculated spectral densities $S(\hbar\omega)$ is close to the ``truth'', taking into account the positions of the peaks and the overall prediction of the vibrational structure. Therefore, we applied a linear scaling factor $\zeta = 1.11$ in $S'(\hbar\omega) = \zeta S(\hbar\omega)$ in all calculations of emission lineshapes to account for the underestimated atomic relaxations.

The Debye--Waller factor (DWF), which quantifies the ratio of the emission contained within the ZPL, was evaluated directly from the calculated luminescence lineshapes. To isolate the ZPL contribution, the spectral region around the ZPL was fitted using a Lorentzian function
\begin{equation}
    \label{eq:lorentzian}
    L(E) = \frac{A}{\pi} 
    \frac{\frac{1}{2}\gamma} {(E - E_{\mathrm{ZPL}})^{2} + (\frac{1}{2}\gamma)^{2}},
\end{equation}
where $A$ is the peak amplitude, and $\gamma$ is the Lorentzian broadening parameter.

\newpage
\section{Embedding methodology}\label{sec:sm_embedding}

\begin{figure}
\includegraphics[width=0.70\textwidth]{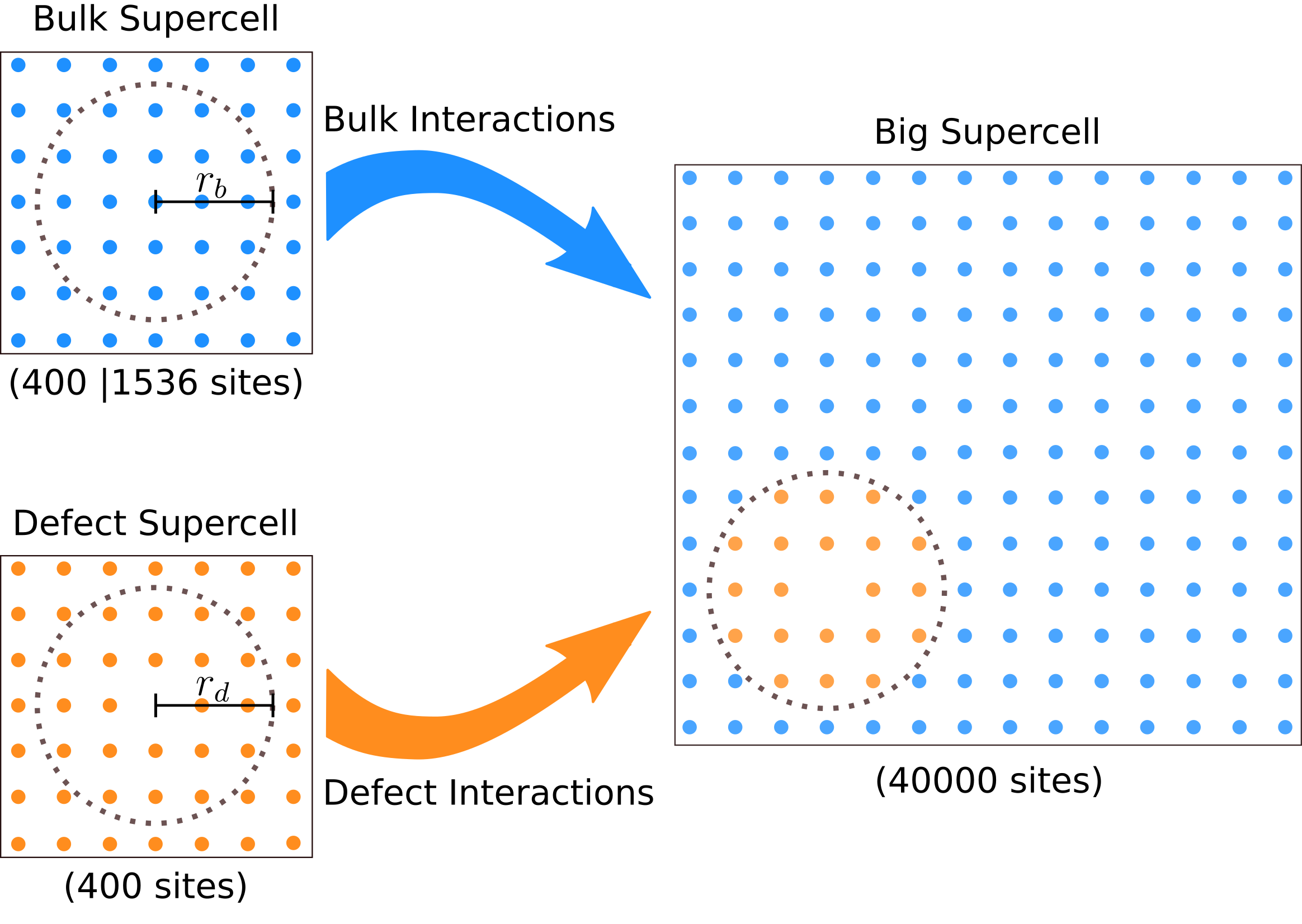}
\caption{Illustration of the embedding methodology for defect vibrational structure calculations.}
\label{fig:embedding}
\end{figure}

The embedding methodology described in Refs.~\cite{Alkauskas_2014,Razinkovas_2021} was employed to model the vibrational properties of large supercells containing a defect. This approach exploits the short-range nature of interatomic forces in semiconductors to construct an effective Hessian matrix for systems containing thousands of atoms. The embedding scheme is illustrated in Fig.~\ref{fig:embedding}. When constructing the Hessian matrix elements, the following criteria are used:
\begin{itemize}
    \item If two atoms lie within a specified cutoff radius $r_{b}$, the corresponding elements from the bulk supercell Hessian matrix are utilized.
    \item For atom pairs within a cutoff radius $r_{d}$ from the defect, matrix elements from the defect-containing supercell are applied.
    \item In all other cases, the matrix elements are set to zero.
\end{itemize}
For the zero strain case, we employed $8\times8\times3$ bulk supercell with 1\,536 atomic sites and $5\times5\times2$ defect supercell containing 400 atomic sites. The corresponding cutoff radii used in the embedding procedure were $r_{b} = 12.200~\text{\r{A}}$ and $r_{d} = 7.690~\text{\r{A}}$. When uniaxial strain of $\pm 2\%$ was applied along the $a$-axis, $5\times5\times2$ supercells consisting of 400 atoms were used for both bulk and defect supercells. For these strained supercells, the bulk cutoff radius was $r_{b} = 7.000~\text{\r{A}}$, while the defect cutoff values were $r_{d} = 7.690~\text{\r{A}}$ for $+2\%$ strain and $r_{d} = 7.536~\text{\r{A}}$ for $-2\%$ strain. 

The Hessian matrix was constructed from a $25\times25\times8$ supercell containing 40\,000 atomic sites. Using the embedding methodology, the vibrational basis is expanded from 1\,194 modes for the $5\times5\times2$ supercell to a total of 119\,994 vibrational modes, corresponding to an increase by a factor of up to $\sim 100$ in the number of modes. This increase in the number of vibrational modes permits the use of smaller Gaussian smoothing widths~$\sigma$, thereby yielding a higher spectral resolution. Such improved resolution is necessary for resolving fine features, for example the splitting of the first phonon sideband peak in the emission lineshape of $\mathrm{V_{Si}^{-}}(k)$. Furthermore, consider $\mathrm{V_{Si}^{-}}(h)$ under $+2\%$ strain: the lowest vibrational mode obtained from a $5\times5\times2$ supercell is 13.5~meV, whereas embedding into a $25\times25\times8$ supercell reduces this value to 3.7~meV. Thus, the embedding methodology provides access to a substantially larger set of vibrational modes, including low-energy acoustic phonons, and a relaxation profile in the dilute limit [see Eq.~\eqref{eq:delta_Q_forces}].

\section{Vibrational structures and localization ratios}\label{sec:sm_IPRs_DOS}

The vibrational properties of defects can be understood by examining the localization of its phonon modes~\cite{Bell_1970}. Bulk-like modes are spatially extended and resemble plane-wave vibrations. In contrast, localized modes are confined to the vicinity of the defect and typically exhibit frequencies outside the host material's phonon band. Between these two cases lie the so-called quasi-local modes~\cite{Alkauskas_2014}, which manifest as defect-induced vibrational resonances embedded within the continuum of bulk phonons. These resonances originate from a collection of vibrational states whose energies fall inside the bulk phonon spectrum. To provide a quantitative measure of phonon localization, we employ the inverse participation ratio (IPR). For a phonon mode $k$, the IPR is defined as~\cite{Bell_1970,Alkauskas_2014}
\begin{equation}
  \label{eq:ipr}
  \mathrm{IPR}_{k}
    = \frac{1}{\sum_{m} \boldsymbol{\eta}_{k;m}^{4}},
\end{equation}
where $\boldsymbol{\eta}_{k;m}$ denotes the three-dimensional, normalized, mass-weighted displacement of atom $m$ in mode $k$. Here,
\begin{equation}
  \boldsymbol{\eta}_{k;m}^{4}
    \equiv \left( \sum_{\alpha} \eta_{k;m\alpha}^{2} \right)^{2},
\end{equation}
with $\eta_{k;m\alpha}$ the mass-weighted displacement of atom $m$ along Cartesian direction $\alpha$. 

The IPR provides an estimate of the number of atoms that participate significantly in a given vibrational mode: $\mathrm{IPR}_{k}=1$ corresponds to a mode localized on a single atom, whereas $\mathrm{IPR}_{k}=N$ indicates that all $N$ atoms in the supercell contribute equally. Intermediate cases, where only $P<N$ atoms vibrate appreciably, yield $\mathrm{IPR}_{k}\approx P$. Although the IPR offers a robust measure of localization, it is often more convenient to introduce the related localization ratio $\beta_{k}$~\cite{Alkauskas_2014}, defined as
\begin{equation}
  \label{eq:beta}
  \beta_{k} = \frac{N}{\mathrm{IPR}_{k}}.
\end{equation}
Here larger values of $\beta_{k}$ indicate stronger localization of phonon mode $k$, making this quantity particularly convenient when comparing defect-induced vibrational characteristics across different supercell sizes.

\begin{figure}
\includegraphics[width=1.00\textwidth]{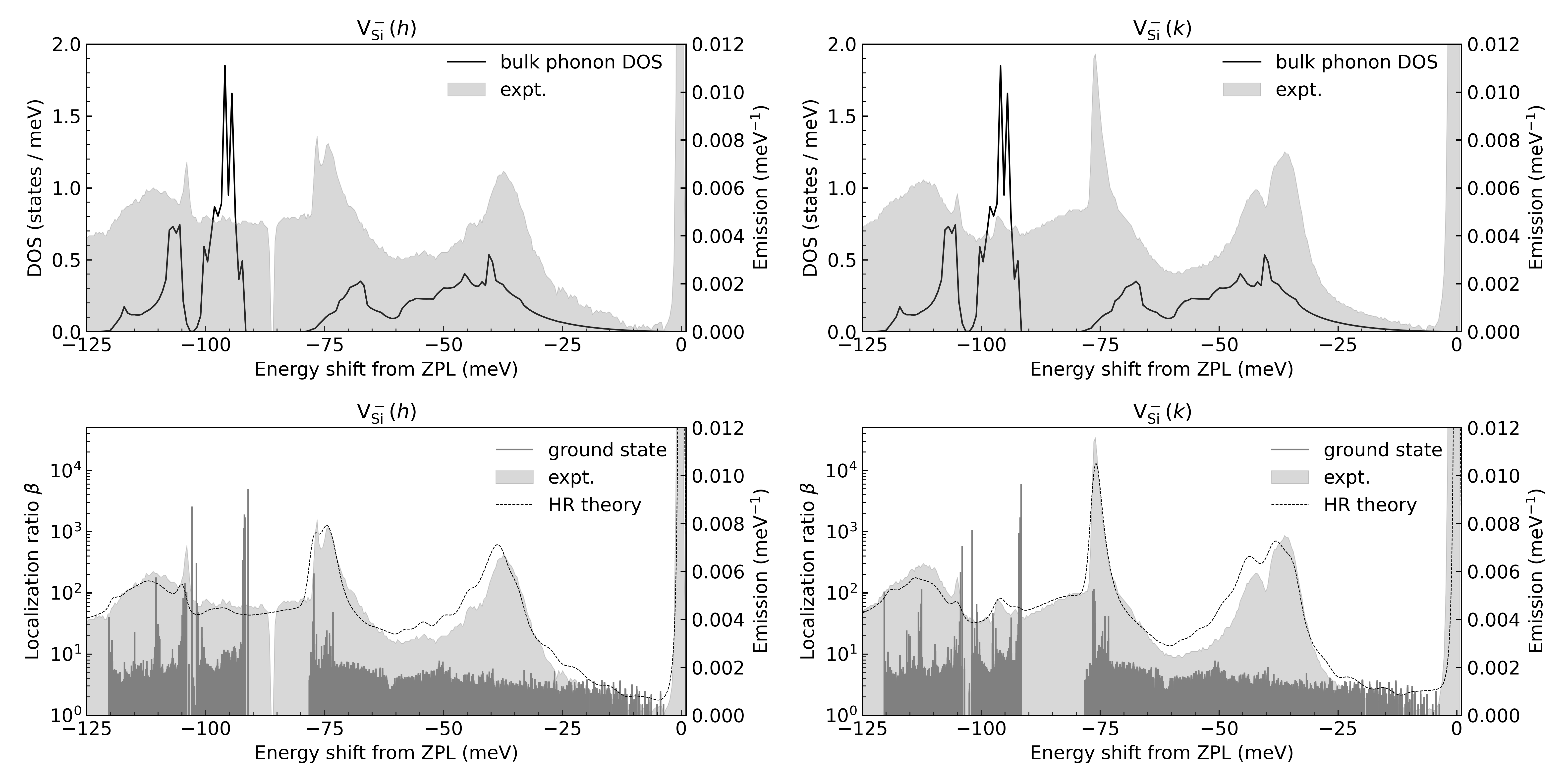}
\caption{Comparison of vibrational and optical properties of the $\mathrm{V_{Si}^{-}}(h)$ and $\mathrm{V_{Si}^{-}}(k)$ defects in unstrained 4H-SiC. Shaded areas in light-gray color are experimental emission lineshapes while dashed gray lines are emission lineshapes calculated using HR theory. The solid black line shows the calculated bulk phonon density of states (DOS). Vertical gray lines show the ground-state vibrational localization ratios $\beta$ computed for a $25\times25\times8$ supercell (40\,000 atomic sites).}
\label{fig:DOS_loc_lum}
\end{figure}

The top panels of Fig.~\ref{fig:DOS_loc_lum} compare the computed bulk phonon density of states (DOS) of unstrained 4H-SiC (solid black line) with the experimental emission lineshapes (shaded light gray area) of the $\mathrm{V_{Si}^{-}}(h)$ and $\mathrm{V_{Si}^{-}}(k)$ defects. In the bottom panels the vertical gray lines are the calculated ground-state vibrational localization ratios $\beta$, obtained using a $25\times25\times8$ supercell containing 40\,000 atomic sites. These localization ratios allow us to identify which types of vibrational modes contribute to the phonon sideband (PSB). The calculated emission lineshapes show that the PSB of the $\mathrm{V_{Si}^{-}}$ defect in 4H-SiC comprises a mixture of bulk-like (BL), quasi-local (QL), and localized vibrational modes together with corresponding replicas (e.g.\ quasi-local replicas, QLR). The peak positions of most prominent PSB features are summarized in Table~\ref{tab:emission_peaks_strain}. For the $\mathrm{V_{Si}^{-}}(k)$ defect, the calculated quasi-local phonon mode at 75.8~meV below the ZPL and its replica at 151.1~meV deviate from the experimentally observed values by 0.2~meV and 0.4~meV, respectively. The spectral resolution of the experimental measurements in this work is approximately 0.25~nm, corresponding to $\sim0.25$--$0.36$~meV in the relevant spectral range. In comparison, the strain-induced shifts under $\pm2\%$ strain along the $a$-axis reach 3.2~meV for the quasi-local mode and 6.6~meV for its replica. These values substantially exceed the experimental spectral resolution, indicating that the predicted strain response should be experimentally resolvable.

\begin{figure}
\includegraphics[width=1.00\textwidth]{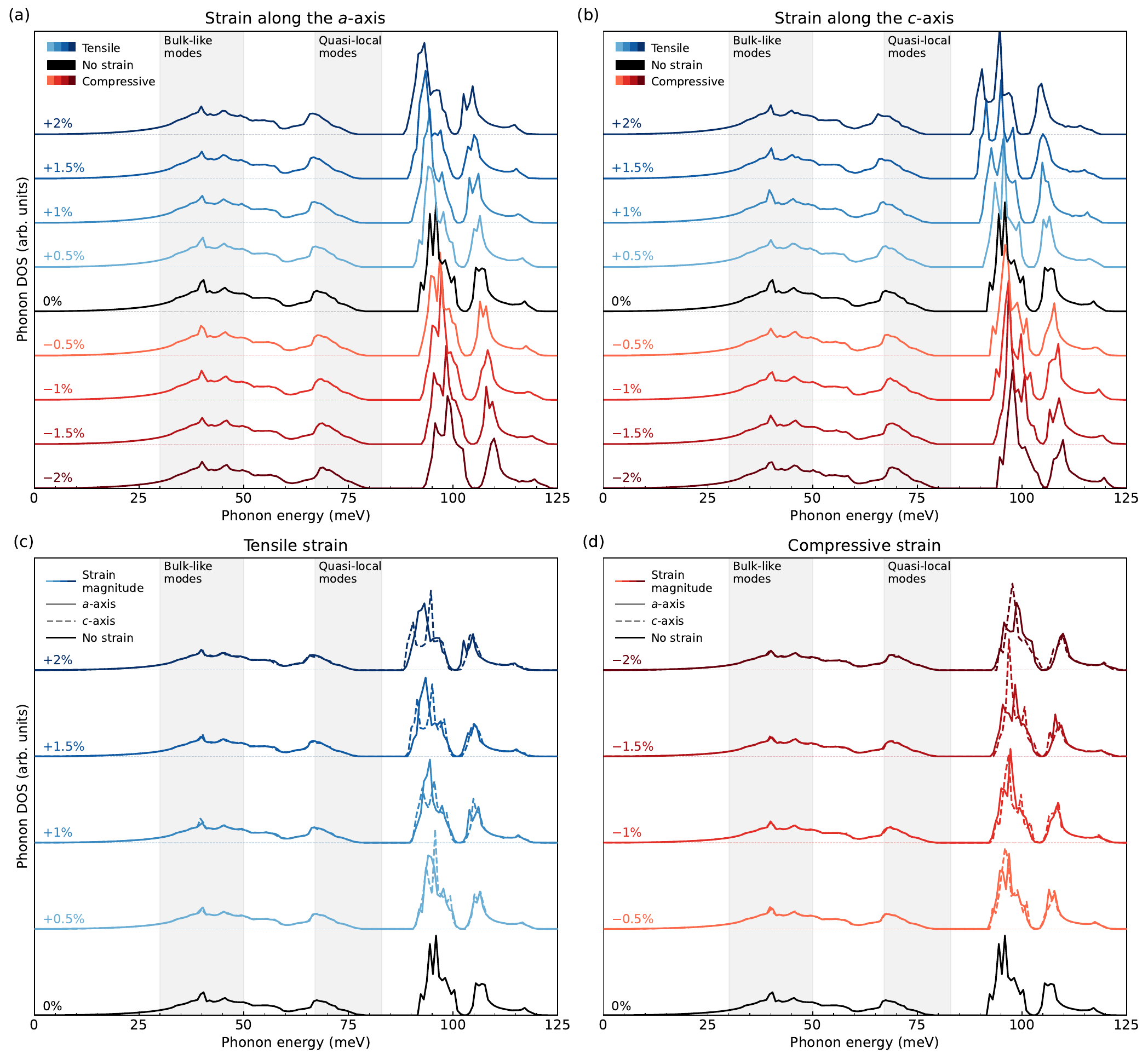}
\caption{Calculated phonon density of states (DOS) of pure 4H-SiC under uniaxial strain. 
Panels (a) and (b) show the evolution of the DOS with strain magnitude along the $a$- and $c$-axes, respectively, with tensile strains (blue) and compressive strains (red) ranging from $\pm0.5\%$ to $\pm2\%$; the unstrained reference is shown in black. 
Panels (c) and (d) compare the $a$-axis (solid lines) and $c$-axis (dashed lines) response at equal strain magnitudes for tensile and compressive strain, respectively. 
Grey shaded regions indicate the frequency ranges of the bulk-like ($30$--$50$~meV) and quasi-local ($67$--$83$~meV) modes of the $\mathrm{V_{Si}^{-}}$ defect for reference.}
\label{fig:DOS}
\end{figure}

Fig.~\ref{fig:DOS} shows how the calculated bulk phonon DOS of 4H-SiC responds to uniaxial strain applied along the $a$- and $c$-axes. The phonon DOS were calculated using $5\times5\times2$ bulk supercells for tensile and compressive strains ranging from $\pm0.5\%$ to $\pm2\%$. Panels~(a) and (b) demonstrate that the low-energy acoustic part of the spectrum (below 60~meV) remains insensitive to strain, whereas the optical phonons in the 90$-$125~meV range exhibit a pronounced strain dependence. For both crystallographic directions, tensile strain (in blue) leads to a systematic ``softening'' of the high-energy optical modes and shifts the dominant peaks toward lower energies, while compressive strain (in red) produces the opposite trend, resulting in mode ``hardening'' of these vibrational modes and shifts toward higher energies. The gradual evolution of the spectra with increasing strain magnitude indicates a monotonic response of both the quasi-local vibrational region ($67$--$83$~meV), relevant to the $\mathrm{V_{Si}^{-}}$ defect, and the high-energy optical phonons to lattice deformation. A comparison between strain applied along the $a$- and $c$-axes is presented in panels~(c) and (d). The strain-induced evolution of the phonon DOS is nearly identical for both crystallographic directions, indicating weak anisotropy in the response of the bulk vibrational spectrum to uniaxial strain.

\begin{figure}
\includegraphics[width=1.00\textwidth]{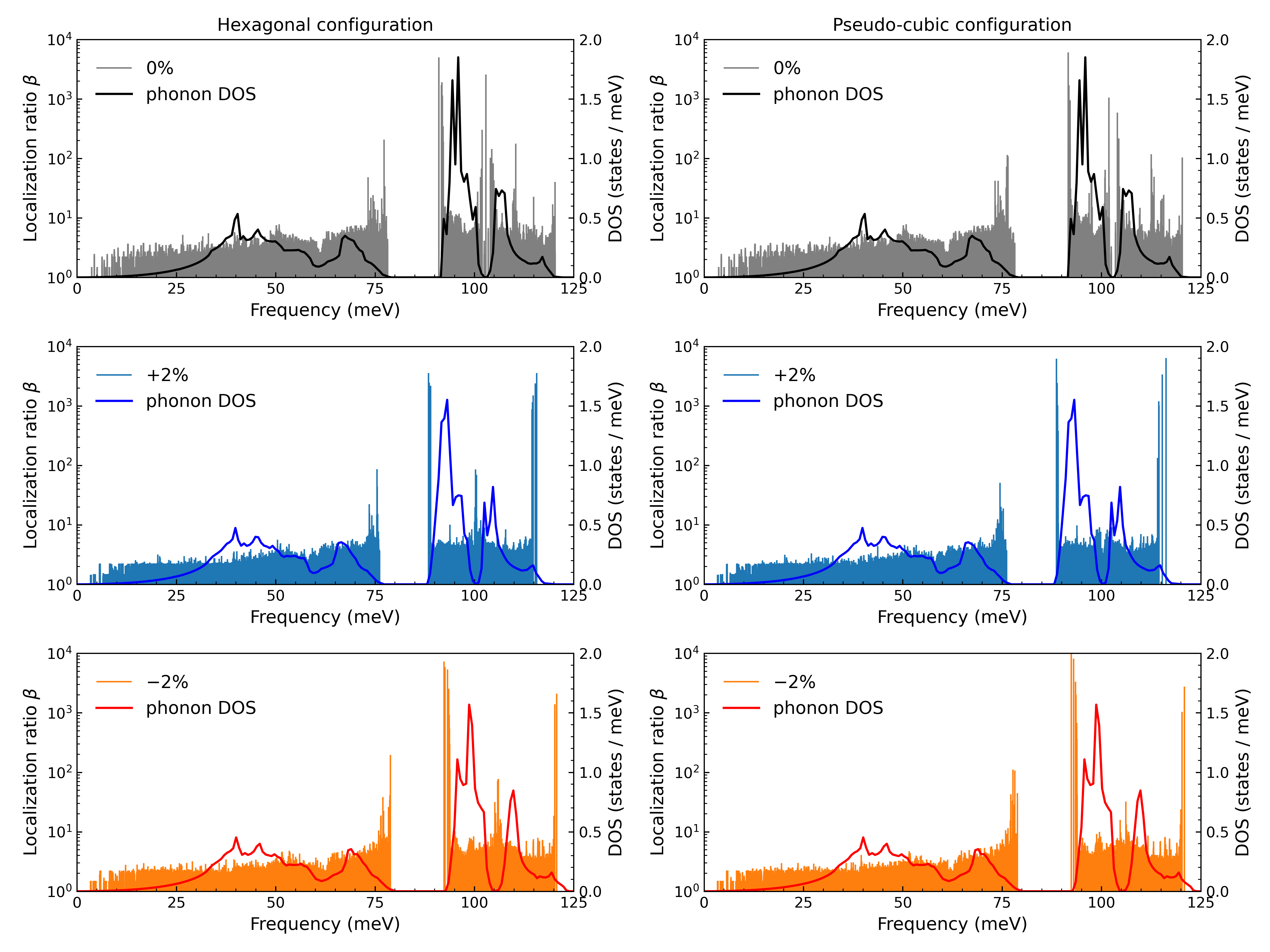}
\caption{Vibrational structures of the $\mathrm{V_{Si}^{-}}(h)$ (left side) and $\mathrm{V_{Si}^{-}}(k)$ (right side) defects at $0\%$, $+2\%$, and $-2\%$ strain along $a$-axis, respectively. Vertical lines correspond to localization ratios $\beta$ of vibrational modes of the $25\times25\times8$ supercell (40\,000 atomic sites) of the ground state. The solid line corresponds to the density of states (DOS) of phonons in pure 4H-SiC at $0\%$ (black line), $+2\%$ (blue line), and $-2\%$ (red line) strain along $a$-axis.}
\label{fig:IPRs}
\end{figure}

Fig.~\ref{fig:IPRs} presents the vibrational structure of the $\mathrm{V_{Si}^{-}}(h)$ (left) and $\mathrm{V_{Si}^{-}}(k)$ (right) configurations under $0\%$, $+2\%$, and $-2\%$ strain along the $a$-axis. Each panel includes the strain-dependent bulk phonon DOS (solid lines) and localization ratios $\beta$ (vertical bars) for each vibrational mode of the negatively charged ground state, computed using a $25\times25\times8$ supercell (40\,000 atomic sites). The overlaid DOS highlights the influence of host crystal vibrational modes how their energies shift under strain.

\begin{table}
    \centering
    \caption{Positions of prominent PSB peaks in the emission lineshape below the ZPL (in meV) for hexagonal and quasicubic configurations under $\pm 2\%$ uniaxial strain along the $a$-axis. Vibrational mode types: BL – bulk-like, QL – quasi-local, QLR – quasi-local replica.}
    \label{tab:emission_peaks_strain}
    \vspace{4pt}
    \begin{tabular}{| c | c  c | c  c | c | c  c | c | c |}
        \hline
        \multirow{3}{*}{Strain} 
        & \multicolumn{5}{c|}{$\mathrm{V_{Si}^{-}}(h)$} 
        & \multicolumn{4}{c|}{$\mathrm{V_{Si}^{-}}(k)$} \\
        \cline{2-10}
        & \multicolumn{2}{c|}{BL} & \multicolumn{2}{c|}{QL} & QLR
        & \multicolumn{2}{c|}{BL} & QL & QLR \\
        \hline
        $0\%$ (expt.) & \multicolumn{2}{c|}{37.4} & 74.2 & 76.5 & 148.0 & 36.2 & 42.0 & 76.0 & 151.5 \\
        $0\%$ (calc.) & \multicolumn{2}{c|}{38.6} & 74.5 & 76.8 & 149.5 & 38.1 & 43.6 & 75.8 & 151.1 \\
        $+2\%$        & 37.3 & 44.5 & \multicolumn{2}{c|}{74.0} & 147.8 & 38.8 & 43.9 & 74.2 & 147.6 \\
        $-2\%$        & \multicolumn{2}{c|}{38.8} & 76.8 & 78.2 & 153.2 & 39.1 & 43.7 & 77.4 & 154.2 \\
        \hline
    \end{tabular}
\end{table}

\section{Temperature-dependent emission lineshapes}

\begin{figure*}
\includegraphics[width=1.00\textwidth]{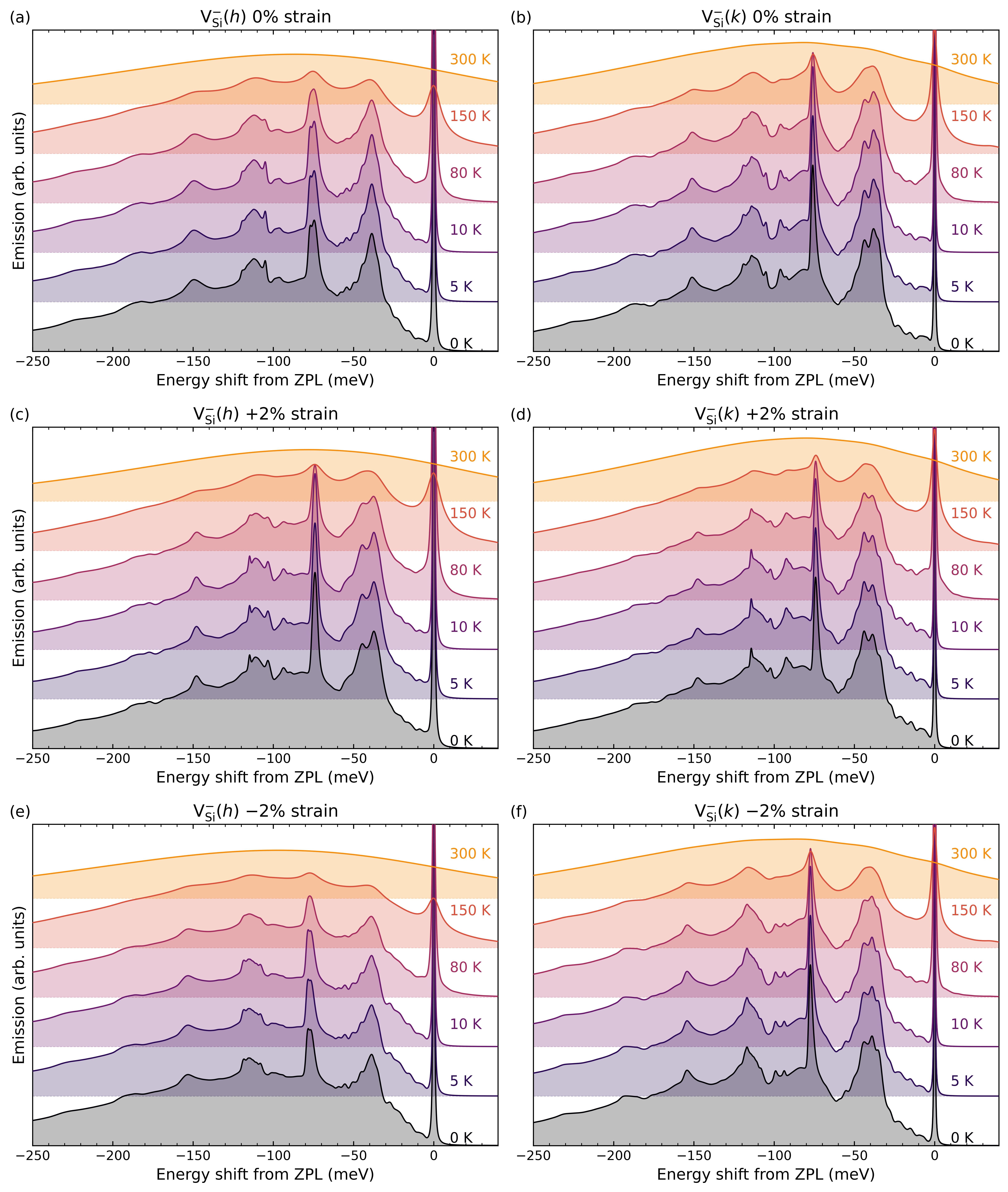}
\caption{Calculated emission lineshapes of the $\mathrm{V_{Si}^{-}}(h)$ and $\mathrm{V_{Si}^{-}}(k)$ defects in 4H-SiC as a function of temperature under uniaxial strain applied along the $a$-axis. 
Panels~(a) and (b) correspond to the zero-strain case, panels~(c) and (d) to $+2\%$ tensile strain, and panels~(e) and (f) to $-2\%$ compressive strain. Lineshapes were computed with r$^{2}$SCAN functional using HR theory and extrapolated to the dilute limit, approximated by a $25\times25\times8$ supercell with 40\,000 atomic sites.}
\label{fig:siv_lum_T}
\end{figure*}

Emission lineshapes calculated at finite temperatures were obtained using the methodology described in Ref.~\cite{Jin_2021}. Within this formalism, the generating function (see Eqs.~\eqref{eq:opt_spectral_func_gen} and~\eqref{eq:gen_func_with_sum}) is extended to include the thermal occupation of phonon modes
\begin{equation}
    \label{eq:gen_func_finite_T}
    G(t,T) = \exp \left\{ 
        -\sum_{k} S_k \left[ \left(1 - e^{i \omega_k t}\right) 
        + 
    \bar{n}_k(T) \left(2 - e^{i \omega_k t} - e^{-i \omega_k t}\right)\right] \right\},
\end{equation}
where $\bar{n}_k(T)$ denotes the average occupation number of the $k$-th phonon mode at temperature $T$, given by the Bose--Einstein distribution
\begin{equation}
    \label{eq:bose_einstein}
    \bar{n}_k(T) = \frac{1}{e^{\hbar \omega_k / k_{\mathrm{B}} T}- 1}.
\end{equation}
The calculated emission lineshapes as a function of temperature are presented in Fig.~\ref{fig:siv_lum_T}. The spectra were evaluated at temperatures of 0, 5, 10, 80, 150, and 300~K. The Lorentzian broadening parameters $\gamma$ (from Eq.~\eqref{eq:opt_spectral_func_gen}) were taken from the experimental measurements reported in Ref.~\cite{Misiara_2025} in order to reproduce the observed temperature dependence of the ZPL linewidths. At cryogenic temperatures  ($T < 30$~K), where acoustic phonon scattering dominates, we employ the model of Ref.~\cite{Udvarhelyi_2020}, whereas at higher temperatures multi-phonon scattering channels become accessible and the broadening is assumed to follow a $T^3$ power law~\cite{Hizhnyakov_1999}.

For the $\mathrm{V_{Si}^{-}}(k)$ defect, all prominent PSB features remain clearly resolved at cryogenic temperatures and persist up to liquid-nitrogen temperatures ($\sim 80$~K). In particular, the sharp feature located 76.0~meV below the ZPL in experiment (75.8~meV in our calculations), associated with quasi-local phonon modes, together with its replica at 151.5~meV (151.1~meV in our calculations), could serve as a robust signature for magnetic-field-free strain sensing. Furthermore, the $\mathrm{V_{Si}^{-}}(k)$ configuration can be selectively addressed optically because its ZPL lies 85.6~meV below that of $\mathrm{V_{Si}^{-}}(h)$ at zero strain. In our calculations, this energy separation varies from 76.9~meV under $-2\%$ compressive strain along the $a$-axis to 98.7~meV under $+2\%$ tensile strain along the $a$-axis. The substantial energetic separation between the two defect configurations therefore makes strain sensing based on the $\mathrm{V_{Si}^{-}}(k)$ defect feasible using standard photoluminescence techniques. However, at room temperature ($300$~K), thermal broadening becomes sufficiently strong that the individual PSB features are no longer spectrally resolved, resulting instead in a broad and featureless PSB.

For the $\mathrm{V_{Si}^{-}}(h)$ defect, the main PSB features are likewise expected to remain distinguishable at cryogenic temperatures and up to liquid-nitrogen temperatures. However, the characteristic double-peak structure associated with quasi-local phonon modes, located 74.0 and 76.4~meV below the ZPL in experiment (74.5 and 76.8~meV in our calculations), is expected to become broadened at elevated liquid-nitrogen temperatures. As a consequence, the corresponding quasi-local phonon replica at 147.9~meV below the ZPL in experiment (149.5~meV in our calculations) is predicted to become less sharp than the analogous replica observed for the $\mathrm{V_{Si}^{-}}(k)$ defect. Since these quasi-local phonon features exhibit the strongest strain-induced spectral shifts, the increased thermal broadening is expected to reduce their suitability for strain sensing compared to the $\mathrm{V_{Si}^{-}}(k)$ configuration. At room temperature, the thermally induced linewidth broadening similarly ends up producing a broad and featureless PSB.

\newpage
\section{Experimental setup}
\label{sm_expt1}
\def\vsih{$\mathrm{V_{Si}^{-}}(h)$}
\def\vsik{$\mathrm{V_{Si}^{-}}(k)$}
\def\DTT{$\mathrm{\Delta T/T}$}

We employ transient absorption (TA) spectroscopy to measure the stimulated emission spectrum of the \vsih\ and \vsik\ cases. We selectively pump at the \vsih\ and \vsik\ ZPLs with a sub-5~cm$\mathrm{^{-1}}$ ($\sim$0.6~meV) narrow-band optical parametric oscillator (OPO) laser (EKSPLA NT242), which allows us to pump the sample anywhere in the 210-2600~nm (0.48-5.9~eV) spectral range. A probe pulse then arrives at the sample about 4~ns later. The measured quantity is the change of the transmission (\DTT) of the probe with and without the presence of the pump. The signal attributed to the stimulated emission is the increase in the transmission of the probe ($\mathrm{\Delta T/T}>0$) on the red side (lower energy) of the spectrum with respect to the ZPL. This is not to be confused with the ground state bleaching (or absorption spectrum), which also results in $\mathrm{\Delta T/T}>0$, but in the blue side (higher energy) of the TA spectrum with respect to the ZPL. In fact, at the ZPL, the signal contains contribution from both the stimulated emission and absorption sides. Combined with the fact that pump scattering for resonant excitation also invalidates the TA spectrum at the ZPL, the stimulated emission spectrum is omitted at this particular spectral position. The TA spectroscopy setup has already been employed in several previous publications~\cite{Luu_2024,Luu_2025,Younesi_2026}. All measurements were performed at sub-10~K temperature.

An advantage of the method employed is the precise pumping of the ZPLs: this allows us to obtain an emission spectrum only from the \vsik\ configuration by pumping resonantly at the ZPL (916~nm/1.353~eV). This is in contrast to a few previous works where the excitation wavelength is fixed (at 730~nm (1.70~eV)~\cite{Udvarhelyi_2020} or 808~nm~\cite{Shang_2020}). However, as the sample studied in this work is in bulk, it contains both \vsih\ and \vsik\ configurations. Since the ZPL of the \vsik\ case is lower in energy compared to the \vsih\ configuration (at 861~nm/1.439~eV), the stimulated emission spectrum of the \vsik\ case is without the contribution of the \vsih. However, the \vsih\ would contain a small contribution of the \vsik\ as the pump also has enough photon energy to excite the \vsik\ configuration. This contribution is substantial at the ZPL of \vsik\ and is removed from the spectrum of the \vsih\ defect. The ZPL of \vsik\ lies 86~meV below the ZPL of \vsih. This ZPL difference falls in the phonon gap (see Fig.~\ref{fig:DOS_loc_lum}) and does not overlap with vibrational features in emission spectra of \vsih\, validating the $\mathrm{V_{Si}^{-}}$ ensemble measurements of stimulated emission.

\begin{figure*}
\includegraphics[width=1.00\textwidth]{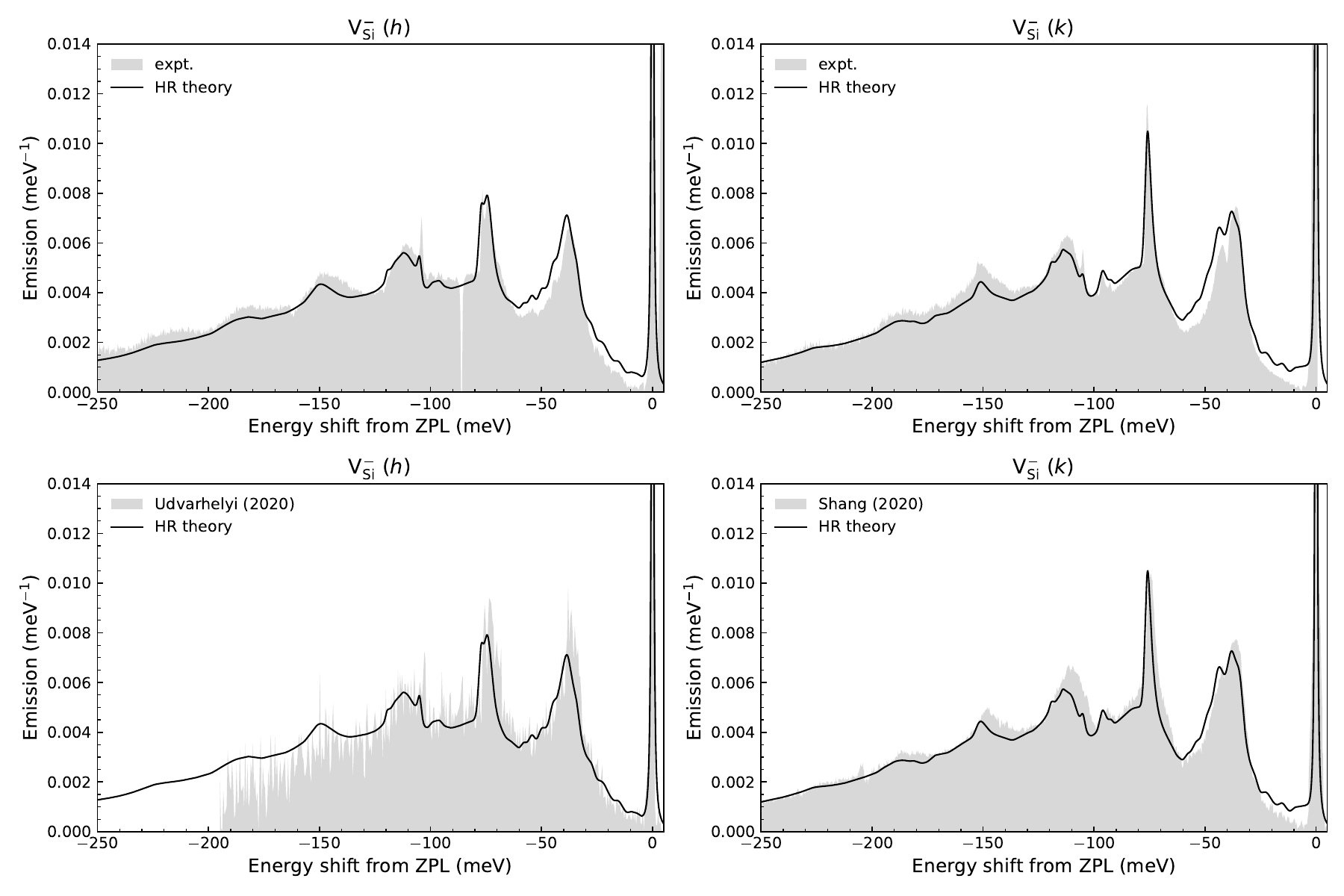}
\caption{Calculated emission lineshapes for supercells with $\mathrm{V_{Si}^{-}}(h)$ and $\mathrm{V_{Si}^{-}}(k)$ defects at zero strain. Lineshapes were computed with r$^{2}$SCAN functional using HR theory and extrapolated to the dilute limit, approximated by a $25\times25\times8$ supercell with 40\,000 atomic sites. Experimental spectra from this work and Refs.~\cite{Udvarhelyi_2020,Shang_2020}.}
\label{fig:lum_abs_strain_other_expt}
\end{figure*}

A comparison between experimental emission lineshapes reported in Refs.~\cite{Udvarhelyi_2020,Shang_2020} and those obtained in this work is shown in Fig.~\ref{fig:lum_abs_strain_other_expt}. For the \vsih\ center, the PSB is more clearly resolved in this work, revealing a distinct double-peak structure associated with quasi-local vibrational modes located at 74.2 and 76.5~meV below the ZPL. In addition, a localized defect vibrational mode at 104.0~meV below the ZPL is clearly resolved. For the \vsik\ center, the emission lineshape obtained in this work more clearly reveals localized vibrational modes at 92.8 and 105.0~meV below the ZPL, as well as a vibrational resonance at 96.5~meV below the ZPL.

\subsection{Oblique incidence probe configuration}
\label{SM_oblique}

 \begin{figure}
    \centering
\includegraphics[width=0.6\textwidth]{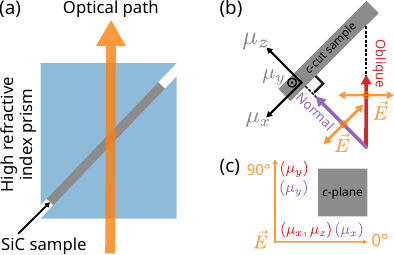}
    \caption{(a) Schematic illustration of the 4H-SiC sample sandwiched between two high refractive index rutile prisms, together with the optical path. (b) Top view schematic of probing geometry for normal (in purple) and oblique (in red) incidence cases of $\mathrm{V_{Si}^{-}}$ electronic transitions with respect to the $c$-plane. The coordinate system defining the transition dipole moments $\mu_x$, $\mu_y$ and $\mu_z$, is indicated. Orange arrows denote the electric field vectors $\vec{E}$ of the probe beam. (c) Front view schematic of the sample ($c$-plane) in the laser beam direction, with summary of the transition dipole moments excited by the $0^\circ$ or $90^\circ$ polarization of the electric field vector.}
    \label{fig:prism_config}
\end{figure}

Under the $C_{3v}$ point group, the transition dipole moments transform as either $(x,y)$ ($E$) or $z$ ($A_1$) in Cartesian coordinates. For example, optical transitions between electronic states $A_2\leftrightarrow\,A_2$ (whose direct product transforms as $A_1$) is allowed by $z$ dipoles, while the $A_2\leftrightarrow\,E$ transition (direct product transforms as $E$) is allowed with $x,y$ dipoles \cite{Udvarhelyi_2020}. We can summarize this experimental scheme with a few useful statements:
\begin{itemize}
    \item 0°-polarization probe can observe all transitions in either \vsih\ or \vsik\ configuration.
    \item 90°-polarization probe only couples to the $\mu_y$ dipole, thus any transition allowed by $\mu_z$ cannot be observed by the probe pulse. 
\end{itemize}
To achieve oblique incidence, a $c$-cut 4H-SiC sample was placed at an angle of 45$^\circ$ with the incident beam. However, due to the considerable refractive index mismatch between the SiC sample and vacuum (approximately 2.7 and 1, respectively), the beam enters the sample at a much smaller angle ($\sim15^\circ$), therefore barely probing the $\mu_z$ component. To resolve this issue, the sample was positioned between two high refractive index rutile right-angle prisms, thereby ensuring the beam passed with minimal refraction (see Fig.~\ref{fig:prism_config} for a schematic representation of the sample configuration). Although passing the laser pulses through the prisms chirps the pulses to a few ps and limits the temporal resolution, the spectra presented here are obtained at a pump-probe delay of about 4~ns.

\end{document}